\documentclass{article}
\usepackage{mathptmx}
\usepackage{amsmath}
\usepackage{amssymb}
\usepackage{hyperref}
\usepackage{graphicx}
\usepackage{ulem}
\usepackage{setspace}
\hypersetup{pdfborder=false}
\title{Integration of the Classical Action for the Quartic Oscillator in 1+1 Dimensions}
\author{Robert L. Anderson{\footnote{\texttt{\href{mailto:andersonr@hal.physast.uga.edu}{andersonr@hal.physast.uga.edu}}}}\\
\vspace{.025in}\\
Department of Physics and Astronomy\\
\vspace{.025in}\\
University of Georgia\\
\vspace{.025in}\\
Athens, Georgia 30602}
\date{\today}
\begin{document}                  

\maketitle

\begin{abstract}
{\noindent}In this paper, we derive an explicit form in terms of end-point data in space-time for the classical action, i.e. integration of the Langrangian along an extremal, for the nonlinear quartic oscillator evaluated on extremals.\end{abstract}

\part{INTRODUCTION}

The action

\begin{align*}
S_{qo}(t_a,y_{a};t_b,y_{b})={\overset{t_b}{\underset{t_{a}}{\int}}}L_{qo}(y{\hspace{3pt}}(t),{\frac{d}{dt}}y{\hspace{3pt}}(t))dt|_{extremal}{\hspace{6pt}},\tag{1.1}    
\end{align*}

where $L_{qo} = {\frac{m}{2}}({\frac{dy}{dt}})^2-{\frac{k_{4}y^4}{4}}$  equals the Lagrangian for the quartic oscillator in 1+1 dimensions, is integrated along an extremal and expressed in terms of the space-time end-point data $(t_a,y_{a}), (t_b,y_{b})$.\\

 We begin in a well-known way by adding and subtracting the kinetic energy to the Lagrangian. Thus we obtain from (1.1), after changing the variable of integration in the remaining integral, the following equivalent expression.

\begin{align*}
S_{qo}(t_a;y;t_b,y)={\overset{y_b}{\underset{y_{a}}{\int}}}m{\hspace{3pt}}{\frac{d}{dt}}y{\hspace{3pt}}dy|_{extremal}{\hspace{6pt}} -E(t_b-t_a)|_{extremal},\tag{1.2}    
\end{align*}

where $E$ is the energy on the extremal (See e.g. Goldstein [1]). Equation (1.2) is the form of the action that we will start from and then derive by integrating the first term in (1.2), which we call the “momentum integral”, thus the desired expression for $S_{qo}$ is obtained. (Some authors call this momentum integral the action.) For our convenience, we refer to the second term in (1.2) as the “energy term”.  The derived action $S_{qo}$ depends only on the end-point data in space-time. \\

In {\bf{Part II Alternative derivation of the Quartic Oscillator Solution}}, we present an approach in which we arrive at the linearization map in [2].  This maps the solutions to Newton’s equations of motion for the quartic oscillator 1-1 onto those of the harmonic oscillator in a way which lends itself to integrating the momentum integral in (1.2).  It involves a parametization of “time” t in terms of an angular coordinate $\theta$  (a cyclic coordinate which takes advantage of the periodic motion of the quartic oscillator and is intrinsic to the harmonic oscillator ho). This results in the time being given by a quadrature involving a known function of $\theta$, as in [2]. As stated in [2] R.C. Santos, J. Santos and J.A.S. Lima [3], first demonstrated the possibility of linearization of the quartic oscillator to the harmonic oscillator. \\

In {\bf{Part III Integration of the momentum integral}}, the results in {\bf{Part II}} lead to an integration of (1.2).  This is a new result and an extension of the results in [3].\\

In {\bf{Part IV Derivation of}} $S_{qo}$, using the results in {\bf{Part II}} and {\bf{Part III}}, we derive a classical action $S_{qo}$ evaluated on an extremal in terms of space-time end-point data and show that Hamilton's equations are satisfied.\\  

In {\bf{Part V Equivalent Actions}}, we present two equivalent actions as variations on the result in {\bf{Part IV}}.  By equivalent we mean they are equal in value on extremals and they produce the same Hamilton's equations.\\

In {\bf{Part VI Conclusion}}, we indicate briefly how the approach in {\bf{Parts III-IV}} can be directly extended to all members of a hierarchy with potential energies $V_{2n}(y_{2n})= {\frac{1}{2n}}k_{2n}y^{2n}_{2n}(t)|_{n{\geq}1}.$  \\

\part{Alternative Derivation of the Quartic Oscillator Solution}

To begin with, we must establish the sign conventions implied by (1.2) for the quartic oscillator 

\begin{align*}
{\overset{y_2}{\underset{y_{1}}{\int}}}m{\hspace{3pt}}{\frac{d}{dt}}y{\hspace{3pt}}(t)dy|_{extremal}{\hspace{6pt}}={\overset{y_2}{\underset{y_{1}}{\int}}}m{\hspace{3pt}}({\pm}){\sqrt{\frac{2E}{m}}}{\hspace{3pt}}{\sqrt{(1-{\frac{V_{qo}(y)}{E}})}}dy|_{extremal}{\hspace{6pt}},\tag{2.1}    
\end{align*}

where ${\frac{V_{qo}(y)}{E}}={\frac{k_4{\hspace{2pt}}y^4}{4E}}$

Taking advantage of the periodicity of any extremal for the quartic oscillator qo, we execute a change of variable to the angular variable $\theta$ by setting

\begin{align*}
\sin^2({\theta}-{\theta}_0)={\frac{V_{qo}(y)}{E}}={\frac{k_4{\hspace{2pt}}y^4}{4E}},\tag{2.2}    
\end{align*}

where $y({\theta}(t_0))=y({\theta}_0)=y(t_0)=y_0=0$ and $dy/dt({\theta}(t_0))={\frac{dy}{dt}}(t_0)={\frac{dy}{dt}}|_{t_0}={\sqrt{2E/m}}$ and $E$ = energy on the extremal.   We have opted not to change the symbol for a function when it depends on a variable through a nested function in order to avoid unnecessarily heavy notation.  Making the signs explicit, (2.1)-(2.2) yield

\begin{align*}
y=({\frac{4E}{k_4}})^{1/4}{\frac{\sin({\theta}-{\theta}_0)}{({\sin}^2({\theta}-{\theta}_0))^{1/4}}}.\tag{2.3}    
\end{align*}

and

\begin{align*}
{\frac{dy}{dt}}={\sqrt{\frac{2E}{m}}}{\hspace{3pt}}{\cos({\theta}-{\theta}_0)}.\tag{2.4}    
\end{align*}

Note, for future use (2.3) implies
\begin{align*}
dy=({\frac{4E}{k_4}})^{1/4}{\frac{1}{2}}{\frac{\cos({\theta}-{\theta}_0)}{({\sin}^2({\theta}-{\theta}_0))^{1/4}}}d{\theta}.\tag{2.5}    
\end{align*}

Now, we are in position to present an alternative derivation of the solution to Newton’s equations of motion (2.7) below for the quartic oscillator.  It involves a parametization of “time” in terms of the angular coordinate.  As we shall see, this results in the time being given by a quadrature involving a known function of $\theta$ . Now differentiating (2.4) yields

\begin{align*}
{\frac{d^2y}{d^2t}}={\sqrt{\frac{2E}{m}}}{\hspace{3pt}}(-{\sin({\theta}-{\theta}_0))}{\frac{d{\theta}}{dt}}.\tag{2.6}    
\end{align*}

Or from Newton’s equation of motion for the quartic oscillator

\begin{align*}
m{\frac{d^2y}{d^2t}}=-k_4y^3,\tag{2.7}    
\end{align*}

we obtain

\begin{align*}
{\frac{ k_4y^3}{m}}={\sqrt{\frac{2E}{m}}}{\hspace{3pt}}{\sin({\theta}-{\theta}_0)}{\frac{d{\theta}}{dt}}.\tag{2.8}    
\end{align*}

Thus, it follows from (2.3) that we obtain the equation that yields $t$ involving $\theta$

\begin{align*}
dt=(k_4E)^{-1/4}m^{1/2}2^{-1}(sin^2({\theta}-{\theta}_0))^{-1/4}d{\theta}.\tag{2.9a}    
\end{align*}

Or, it’s integrated form which yields $t$ (in quadrature) involving a known function of $\theta$

\begin{align*}
t({\theta})-t_0={\overset{\theta}{\underset{{\theta}_0}{\int}}}(k_4E)^{-1/4}{\frac{m^{1/2}}{2}}(sin^2({\theta}'-{\theta}_0))^{-1/4}d{\theta}'.\tag{2.9b}    
\end{align*}

The inverse of (2.9a) is given by 

\begin{align*}
d{\theta}=({\frac{2k_4}{m}})^{1/2}(y^2)^{1/2}dt,\tag{2.10a}    
\end{align*}

and it’s integrated form is given by

\begin{align*}
{\theta}(t)-{\theta}_0={\overset{t}{\underset{t_0}{\int}}}({\frac{2k_4}{m}})^{1/2}(y^2(t'))^{1/2}dt',\tag{2.10b}    
\end{align*}

where the integration is along an extremal.

The equivalence to the linearization  map  given in [2]  is specified by setting ${\theta}-{\theta}_0={\omega}({\hat{t}}-{\hat{t}}_0)$, where $k_2=m{\omega}^2=spring$ constant of the harmonic oscillator $ho$ and $\hat{t}$ equals the time of the $ho$ corresponding to $t$ of the $qo$.\\

Then (2.9b) and (2.10b)  are equivalent to one half of the linearization map in [2]. The other half of the linearization map is given by 

\begin{align*}
{\frac{(y^2(t))^{1/2}y(t)}{(4E/k_4)^{1/2}}}={\sin}({\theta}-{\theta}_0)={\frac{x_{ho}({\hat{t}})}{(2E/k_2)^{1/2}}}.\tag{2.11}    
\end{align*}

Equation (2.2) plus equation (2.4) imply

\begin{align*}
{\frac{4E}{k_4}}=[y{^4_b}+y{^4_a}-2(y{^2_b})^{1/2}y_b(y{^2_a})^{1/2}y_a{\hspace{3pt}}{\cos}({\theta}_b-{\theta}_a)]/{\sin}^2({\theta}_b-{\theta}_a)\tag{2.12}    
\end{align*}

where $({\theta}_a-{\theta}_a)$ is given by (2.10b).\\

Finally, in this paragraph, given the end-point data how does one determine all other quantities.\\

One is given $(y_a, t_a)$ and $(y_b, t_b)$ on an $qo$ extremal.  The linearization map yields $x_a$ and $x_b$ on the corresponding $ho$ extremal as well as $E_{ho}=E_{qo}=E$.  This implies from (2.12) the $ho$ time differences ($\hat{t}_a-\hat{t}_b$) and ($\hat{t}_b-\hat{t}_o$), where $\hat{ }$ refers to $ho$ times, are known.  Now we can set $t_o=\hat{t}_o$.  \\                                                                                                                                                                                                                                                                                                                                                                                                                                               

From [4], as a result of mapping extremals for the $ho$ $1-1$ onto extremals to the $qo$, we have from [4],

\begin{align*}
{\sin}({\theta}(t)-{\theta}_0)=(4E/k_4)^{-1/2}{\Biggl{[}}{\frac{(y^{2}_{b})^{1/2}y_b{\sin}({\theta}(t)-{\theta}_a)+(y^{2}_{a})^{1/2}y_a{\sin}({\theta}_b-{\theta}(t))}{{\sin}({\theta}_b-{\theta}_a)}}{\Biggr{]}},\tag{2.13}
\end{align*}

and

\begin{align*}
&{\cos}({\theta}(t)-{\theta}_0)\\
&=(4E/k_4)^{-1/2}{\Biggl{[}}{\frac{(y^{2}_{b})^{1/2}y_b{\cos}({\theta}(t)-{\theta}_a)-(y^{2}_{a})^{1/2}y_a{\cos}({\theta}_b-{\theta}(t))}{{\sin}({\theta}_b-{\theta}_a)}}{\Biggr{]}},\tag{2.14}
\end{align*}

Now (2.13) and (2.14) imply e.g.

$tan({\theta}_b-{\theta}_0)={\frac{(y^{2}_{b})^{1/2}y_b{\sin}({\theta}_b-{\theta}_a)}{(y^{2}_{b})^{1/2}y_b{\cos}({\theta}_b-{\theta}_a)-(y^{2}_{a})^{1/2}y_a}}$

where $\hat{t}_b-\hat{t}_o$ - $(\hat{t}_a-\hat{t}_o)$ = $\hat{t}_b-\hat{t}_a$ and $\omega\hat{t}={\theta}(t)$ yields ${\theta}_0$.\\

Everything else follows from the development in {\bf{Part III}}.

\part{Integration of ${\overset{y_b}{\underset{y_a}{\int}}}m{\hspace{3pt}}{\frac{dy_{qo}}{dt}}dy_{qo}|_{extremal}$}

The problem of integrating (1.2) is the problem of integrating (2.1). Therefore, using (2.2) , (2.4) ,and (2.5), we obtain

\begin{align*}
&{\overset{y_b}{\underset{y_a}{\int}}}m{\hspace{3pt}}{\frac{d}{dt}}y{\hspace{3pt}}(t)dy|_{extremal}{\hspace{6pt}}={\overset{y_b}{\underset{y_a}{\int}}}m{\hspace{3pt}}({\pm}){\sqrt{\frac{2E}{m}}}{\hspace{3pt}}{\sqrt{(1-{\frac{k_4y^4}{4E}})}}dy|_{extremal}\\
&={\overset{{\theta}_b}{\underset{{\theta}_a}{\int}}}m{\hspace{3pt}}{\sqrt{\frac{2E}{m}}}{\cos}({\theta}'-{\theta}_0)({\frac{4E}{k_4}})^{1/4}({\frac{1}{2}})({\sin}^2({\theta}'-{\theta}_0))^{-1/4}{\cos}({\theta}'-{\theta}_0)d{\theta}'\tag{3.1}    
\end{align*}

Effecting the integration by parts, where ${\frac{d}{d{\theta}}}fg={\frac{df}{d{\theta}}}g+f{\frac{dg}{d{\theta}}},f=({\sin}^2({\theta}-{\theta}_0))^{3/4}$  and $g={\frac{2{\cos}({\theta}-{\theta}_0)}{3{\sin}({\theta}-{\theta}_0)}}$

yields

\begin{align*}
&{\overset{y_b}{\underset{y_a}{\int}}}m{\hspace{3pt}}{\frac{dy_{qo}}{dt}}dy_{qo}|_{extremal}=m{\sqrt{\frac{2E}{m}}}{\hspace{3pt}}({\frac{4E}{k_4}})^{1/4}({\frac{1}{2}})({\frac{2}{3}}){\hspace{4pt}}[{\hspace{4pt}}({\frac{2}{m^{1/2}}}(k_4E)^{1/4})({\frac{m^{1/2}}{2}}{\frac{1}{(k_4E)^{1/4}}}){\overset{{\theta}_b}{\underset{{\theta}_a}{\int}}}(sin^2({\theta}-{\theta}_0))^{-1/4}d{\theta}\\
&+(sin^2({\theta}-{\theta}_0))^{3/4}{\frac{{\cos}({\theta}-{\theta}_0)}{{\sin}({\theta}-{\theta}_0)}}|^{{\theta}_b}_{{\theta}_a}{\hspace{4pt}}]\\
&={\frac{4E}{3}}(t_2-t_1)+{\frac{2m^{1/2}(E)^{3/4}}{3(k_4)^{1/4}}}(sin^2({\theta}-{\theta}_1))^{3/4}{\frac{{\cos}({\theta}-{\theta}_0)}{{\sin}({\theta}-{\theta}_0)}}|^{{\theta}_b}_{{\theta}_a}{\rbrace}\tag{3.2}    
\end{align*}

Finally, from (2.9b), we have

\begin{align*}
&{\overset{y_b}{\underset{y_a}{\int}}}m{\hspace{3pt}}{\frac{dy_{qo}}{dt}}dy_{qo}|_{extremal}={\frac{4E}{3}}(t_b-t_a)+{\frac{2m^{1/2}(E)^{3/4}}{3(k_4)^{1/4}}}(sin^2({\theta}-{\theta}_0))^{3/4}{\frac{{\cos}({\theta}-{\theta}_0)}{{\sin}({\theta}-{\theta}_0)}}|^{{\theta}_b}_{{\theta}_a}\\
&={\frac{4E}{3}}(t_b-t_a)+{\frac{1}{3}}({\frac{mk_4}{2}})^{1/2}(y^2(t))^{3/2}{\frac{{\cos}({\theta}-{\theta}_0)}{{\sin}({\theta}-{\theta}_0)}}|^{{\theta}_b}_{{\theta}_a},\tag{3.3}    
\end{align*}

where ${\theta}-{\theta}_0$ is given by (2.10b).

\part{Determination of an $S_{qo}$.}

The developments in {\bf{Part II}} and {\bf{Part III}} lead directly to the following determination of $S_{qo}$.

It follows from (3.3) that (1.2) is given by

\begin{align*}
&S_{qo}(t_a;y;t_b,y_b)={\overset{y_b}{\underset{y_{a}}{\int}}}m{\hspace{3pt}}{\frac{d}{dt}}y{\hspace{3pt}}dy|_{extremal}{\hspace{6pt}} -E(t_b-t_a)|_{extremal}\\ 
&={\frac{4E}{3}}(t_b-t_a)+{\frac{1}{3}}({\frac{mk_4}{2}})^{1/2}(y^2(t))^{3/2}{\frac{{\cos}({\theta}-{\theta}_0)}{{\sin}({\theta}-{\theta}_0)}}|^{{\theta}_b}_{{\theta}_a}-E(t_b-t_a)\\
&={\frac{1}{3}}({\frac{mk_4}{2}})^{1/2}{\hspace{4pt}}[{\hspace{4pt}}(y^2_b)^{3/2}{\frac{{\cos}({\theta}_b-{\theta}_0)}{{\sin}({\theta}_b-{\theta}_0)}}-(y^2_a)^{3/2}{\frac{{\cos}({\theta}_a-{\theta}_0)}{{\sin}({\theta}_a-{\theta}_0)}}{\hspace{4pt}}]{\hspace{4pt}}+{\frac{E}{3}}(t_b-t_a)\tag{4.1}  
\end{align*}

Therefore, using (2.10b), we obtain

\begin{align*}
&S_{qo}(t_a;y;t_b,y_b)=\\
&={\frac{1}{3}}({\frac{mk_4}{2}})^{1/2}{\hspace{4pt}}[{\hspace{4pt}}(y^2_b)^{3/2}{\frac{{\cos}{\overset{t_b}{\underset{t_0}{\int}}}({\frac{2k_4}{m}})^{1/2}(y^2(t'))^{1/2}dt'}{{\sin}\overset{t_b}{\underset{t_0}{\int}}({\frac{2k_4}{m}})^{1/2}(y^2(t'))^{1/2}dt'}}-(y^2_a)^{3/2}{\frac{{\cos}{\overset{t_{a}}{\underset{t_o}{\int}}}({\frac{2k_4}{m}})^{1/2}(y^2(t'))^{1/2}dt'}{{\sin}\overset{t_{a}}{\underset{t_o}{\int}}({\frac{2k_4}{m}})^{1/2}(y^2(t'))^{1/2}dt'}}{\hspace{4pt}}]\\
&+{\frac{E}{3}}(t_b-t_a)\tag{4.2}  
\end{align*}

This is expressed in the endpoint variables as required.  This implies

\begin{align*}
&{\frac{{\partial}S_{qo}}{{\partial}{y_b}}}=p_{{qo}_b}=({\frac{mk_4}{2}})^{1/2}{\hspace{3pt}}(y^2_b)^{1/2}{\hspace{3pt}}y_b{\frac{{\cos}{\overset{t_b}{\underset{t_0}{\int}}}({\frac{2k_4}{m}})^{1/2}(y^2(t'))^{1/2}dt'}{{\sin}\overset{t_b}{\underset{t_0}{\int}}({\frac{2k_4}{m}})^{1/2}(y^2(t'))^{1/2}dt'}}=({\frac{mk_4}{2}})^{1/2}({\frac{4E}{k_4}})^{1/2}{\cos}{\overset{t_b}{\underset{t_0}{\int}}}({\frac{2k_4}{m}})^{1/2}(y^2(t'))^{1/2}dt'\\
&=(2mE)^{1/2}{\cos}{\overset{t_b}{\underset{t_0}{\int}}}({\frac{2k_4}{m}})^{1/2}(y^2(t'))^{1/2}dt',\\
&{\frac{{\partial}S_{qo}}{{\partial}{t_b}}}={\frac{1}{3}}({\frac{mk_4}{2}})^{1/2}{\hspace{4pt}}[{\hspace{4pt}}(y^2_b)^{3/2}{\hspace{3pt}}{\frac{(-1)}{{\sin}^2\overset{t_b}{\underset{t_0}{\int}}({\frac{2k_4}{m}})^{1/2}(y^2(t'))^{1/2}dt'}}{\hspace{4pt}}]{\hspace{4pt}}({\frac{2k_4}{m}})^{1/2}(y^2_b)^{1/2}+{\frac{1}{3}}E=-E.\tag{4.3}  
\end{align*}

After using (2.11) this checks with $m$ times (2.4) for $^{p_{{qo}_b}}$ and $^{{\partial}/{\partial}t_b}$ obviously checks.\\

The $a$-differentiations parallel the $b$-differentiations and yield

\begin{align*}
&{\frac{{\partial}S_{qo}}{{\partial}{y_a}}}=-p_{{qo}_a}=-({\frac{mk_4}{2}})^{1/2}{\hspace{3pt}}(y^2_a)^{1/2}{\hspace{3pt}}y_a{\frac{{\cos}{\overset{t_o}{\underset{t_a}{\int}}}({\frac{2k_4}{m}})^{1/2}(y^2(t'))^{1/2}dt'}{{\sin}\overset{t_a}{\underset{t_0}{\int}}({\frac{2k_4}{m}})^{1/2}(y^2(t'))^{1/2}dt'}}=-({\frac{mk_4}{2}})^{1/2}({\frac{4E}{k_4}})^{1/2}{\cos}{\overset{t_0}{\underset{t_a}{\int}}}({\frac{2k_4}{m}})^{1/2}(y^2(t'))^{1/2}dt'\\
&=-(2mE)^{1/2}{\cos}{\overset{t_0}{\underset{t_a}{\int}}}({\frac{2k_4}{m}})^{1/2}(y^2(t'))^{1/2}dt',\\
&{\frac{{\partial}S_{qo}}{{\partial}{t_a}}}=-{\frac{1}{3}}({\frac{mk_4}{2}})^{1/2}{\hspace{4pt}}[{\hspace{4pt}}(y^2_b)^{3/2}{\hspace{3pt}}{\frac{(-1)}{{\sin}^2\overset{t_o}{\underset{t_a}{\int}}({\frac{2k_4}{m}})^{1/2}(y^2(t'))^{1/2}dt'}}{\hspace{4pt}}]{\hspace{4pt}}({\frac{2k_4}{m}})^{1/2}(y^2_b)^{1/2}-{\frac{1}{3}}E=+E.\tag{4.4}  
\end{align*}

\part{Equivalent Actions}

Here, we present two examples of equivalent actions as variations on this result. By equivalent we mean they are equal in value on extremals and they both produce the same Hamilton equations.\\

{\vspace{50pt}}

{\bf{First Variation}}:

This variation follows from the indentities

\begin{align*}
&{\sin}({\theta}-{\theta}_0)={\sin}(\theta{\pm}{\theta}_{\max}-{\theta}_0)={\cos}({\theta}-{\theta}_{\max}),\\
&{\cos}({\theta}-{\theta}_0)={\cos}(\theta{\pm}{\theta}_{\max}-{\theta}_0)=-{\sin}({\theta}-{\theta}_{\max}),\tag{5.1}
\end{align*}

which implies that (4.2) transforms to the expression

\begin{align*}
&S_{qo}(t_a,y_a;t_b,y_b)={\overset{y_b}{\underset{y_{a}}{\int}}}m{\hspace{3pt}}{\frac{d}{dt}}y{\hspace{3pt}}dy|_{extremal}{\hspace{6pt}} -E(t_b-t_a)|_{extremal}\\ 
&={\frac{1}{3}}({\frac{mk_4}{2}})^{1/2}{\hspace{4pt}}[{\hspace{4pt}}-(y^2_b)^{3/2}{\frac{{\sin}{\overset{t_b}{\underset{t_{\max}}{\int}}}({\frac{2k_4}{m}})^{1/2}(y^2(t'))^{1/2}dt'}{{\cos}\overset{t_b}{\underset{t_{\max}}{\int}}({\frac{2k_4}{m}})^{1/2}(y^2(t'))^{1/2}dt'}}+(y^2_a)^{3/2}{\frac{{\sin}{\overset{t_{a}}{\underset{t_{\max}}{\int}}}({\frac{2k_4}{m}})^{1/2}(y^2(t'))^{1/2}dt')}{{\cos}\overset{t_{a}}{\underset{t_{\max}}{\int}}({\frac{2k_4}{m}})^{1/2}(y^2(t'))^{1/2}dt')}}{\hspace{4pt}}]+{\frac{E}{3}}(t_b-t_a).\tag{5.2}
\end{align*}

{\bf{Second Variation}}:

Eq. 5.2 is equivalent to
\begin{align*}
S_{qo}(t_a,y_a;t_b,y_b)= \\
{\frac{1}{3}}({\frac{mk_4}{2}})^{1/2}{\hspace{4pt}}{\Bigg{\lbrace}}[{\hspace{1pt}}(y^2_b)^{3/2}{\cos}{\overset{t_b}{\underset{t_{\max}}{\int}}}({\frac{2k_4}{m}})^{1/2}(y^2(t'))^{1/2}dt'-(y^2_b)^{3/2}/{\cos}{\overset{t_b}{\underset{t_{\max}}{\int}}({\frac{2k_4}{m}})^{1/2}(y^2(t'))^{1/2}dt']}/{{\sin}\overset{t_b}{\underset{t_{\max}}{\int}}({\frac{2k_4}{m}})^{1/2}(y^2(t'))^{1/2}dt'}\\
-[(y^2_a)^{3/2}{\cos}{\overset{t_{\max}}{\underset{t_{a}}{\int}}}({\frac{2k_4}{m}})^{1/2}(y^2(t'))^{1/2}dt'-(y^2_a)^{3/2}/{\cos}{\overset{t_{\max}}{\underset{t_{a}}{\int}}({\frac{2k_4}{m}})^{1/2}(y^2(t'))^{1/2}dt']/{{\sin}\overset{t_{\max}}{\underset{t_{a}}{\int}}({\frac{2k_4}{m}})^{1/2}(y^2(t'))^{1/2}dt')}}{\hspace{4pt}}]{\Bigg{\rbrace}}\\
+{\frac{1}{3}}E(t_b-t_a).\tag{5.3}
\end{align*}

Comment: The signs and the limits of integration have to be carefully watched in these calculations.

The identity
\begin{align*}
-(y^2_b)^{3/2}/{\cos}({\theta}_b-{\theta}_{\max})=-3y_by_{\max}(y^2_{\max})^{1/2}+2(y^2_{\max})^{3/2}({\cos}^2({\theta}_b-{\theta}_{\max}))^{3/4}/{\cos}({\theta}_b-{\theta}_{\max}),\tag{5.4}
\end{align*}

follows from
$(y^2_b)^{1/2}y_b=(y^{2}_{\max})^{1/2}{\hspace{3pt}}y_{\max}{\hspace{3pt}}{\cos}({\theta}_b-{\theta}_{\max}).$

Similarly for the $a$ endpoint, thus we obtain the result reported in [4].\\

The results given in [4] were obtained before the integration result reported here in Part IV was obtained.\\

{\vspace{-10pt}}

\part{Conclusion}

One can parallel the development in {\bf{Parts III - IV}} for an hierarchy with potential energies

\begin{align*}
V_{2n}(y)= {\frac{1}{2n}}k_{2n}y^{2n}|_{n{\geq}1}\tag{6.1}
\end{align*}

Starting with setting

\begin{align*}
\sin^2({\theta}-{\theta}_0)={\frac{V_{2n}(y)}{E}}={\frac{k_{2n}{\hspace{2pt}}y^{2n}}{2nE}},\tag{6.2}    
\end{align*}

one can parallel {\bf{Part III}}.\\

Then integration by parts in these cases is effected by ${\frac{d}{d{\theta}}}fg={\frac{df}{d{\theta}}}g+f{\frac{dg}{d{\theta}}},f=({\sin}^2({\theta}-{\theta}_0)^{(n+1)/2n}$  and $g={\frac{n{\cos}({\theta}-{\theta}_0)}{(n+1){\sin}({\theta}-{\theta}_0)}}$.\\

This then parallels the development in {\bf{Part IV}}.\\

The linearization map for these cases is given in [2].

\section*{References}

[1] H. Goldstein, {\textit{Classical Mechanics}}, (Addison-Wesley Publishing Company, Reading, Mass. 1980).

[2] Robert L. Anderson, “An Invertible Linearization Map for the Quartic Oscillator”, JMP {\textbf{51}} , 122904 (2010).

[3] R. C. Santos, J. Santos and J. A. S. Lima, ``Hamilton-Jacobi approach for power-law potentials," Braz. J. Phys. {\textbf{36}} (0.4A), 2006 pp.1257-1261.

[4] Robert L. Anderson, “Actions for an Hierarchy of Attractive Nonlinear Oscillators Including the Quartic Oscillator in 1+1 Dimensions, arXiv.org:1204.0768.

\end{document}